\documentclass{article} 
\usepackage{nips15submit_e,times}
\usepackage{hyperref}
\usepackage{url}
\usepackage{pdfpages}

\usepackage{nips15submit_e,times}
\usepackage{graphicx}
\usepackage{latexsym}
\usepackage{balance}  
\usepackage{amsmath}
\usepackage{algorithm}
\usepackage[noend]{algpseudocode}
\usepackage{tikz}
\usepackage{subcaption}
\usepackage{graphicx}
\usepackage{float}
\usetikzlibrary{arrows,shapes, backgrounds}

\newtheorem{defn}{Definition}
\newtheorem{thm}{Theorem}

\long\def\comment#1{}

\makeatletter
\def\BState{\State\hskip-\ALG@thistlm}
\makeatother

\usepackage{subcaption}
\DeclareCaptionSubType*{algorithm}

\DeclareCaptionLabelFormat{alglabel}{Alg.~#2}

\title{Parameter Database : Data-centric Synchronization for Scalable Machine Learning}
\author{
Naman Goel, Divyakant Agrawal, Sanjay Chawla, Ahmed Elmagarmid\\
\texttt{\{ngoel, dagrawal, schawla, aelmagarmid\}@qf.org.qa} \\
}

\nipsfinalcopy 

\begin{document}
\includepdf[pages={1}]{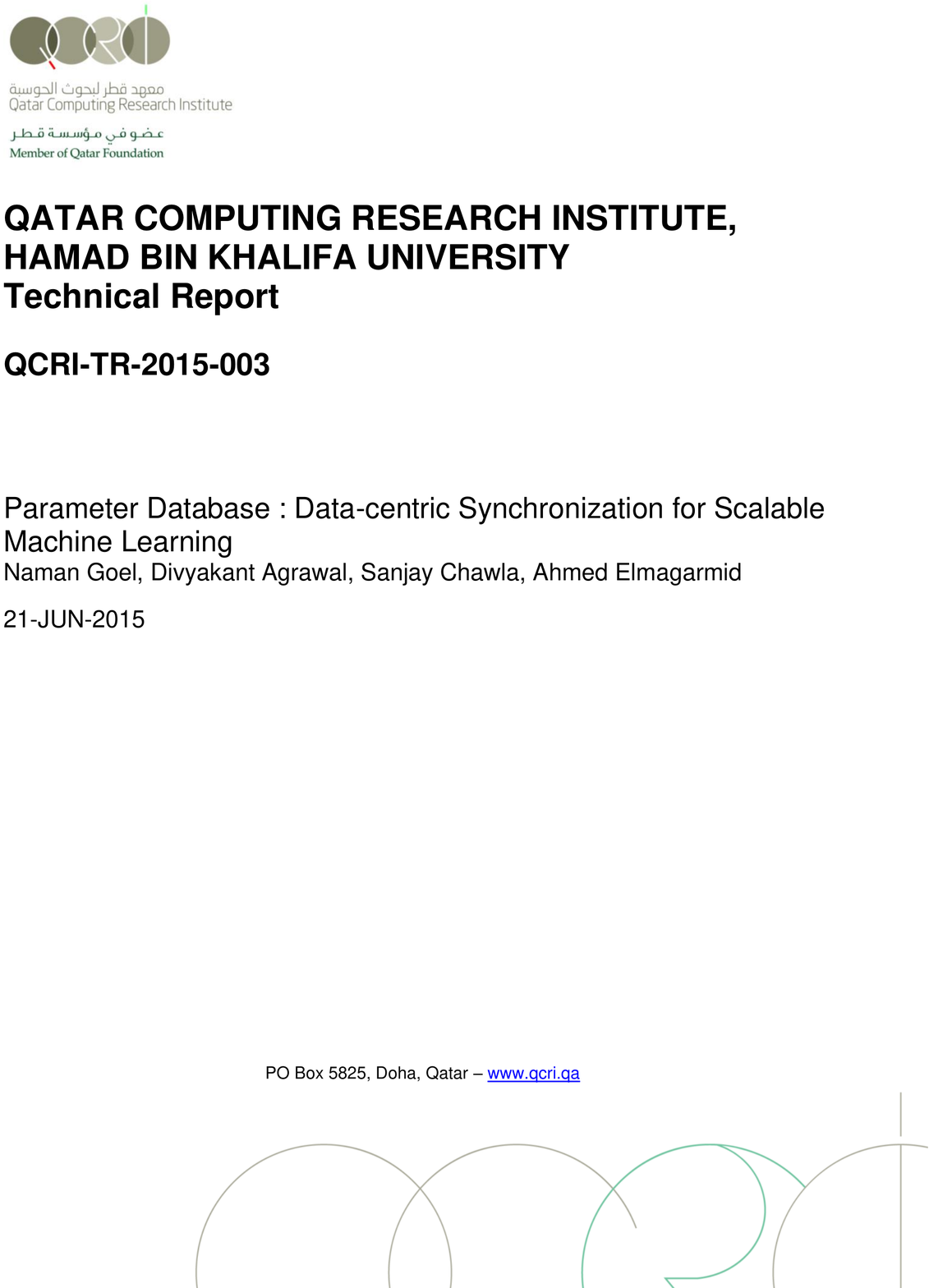}
\maketitle

\begin{abstract}
	We propose a new data-centric synchronization framework for carrying out of machine learning (ML) 
	tasks in a distributed environment. Our framework exploits the iterative nature of 
	ML algorithms and relaxes the application agnostic bulk synchronization parallel (BSP)  
	paradigm that has previously been used for distributed machine learning. Data-centric synchronization
	complements function-centric synchronization based on using stale updates to increase
	the throughput of distributed ML computations.  Experiments to validate our framework
	suggest that we can attain substantial improvement over BSP while guaranteeing sequential
	correctness of ML tasks.
\end{abstract}

\section{Introduction and Related Work}
In an increasing number of application domains ranging from to speech and image recognition systems to online advertising, both the size of data sets and the complexity of learning models continues to increase. It is now not uncommon to train machine
learning models which consist of over a billion
parameters\footnote{the terms ``parameter'' and ``feature'' are often used interchangeably.}  and data
points~\cite{li_pserver,dean2012nips}.  For such large scale machine
learning models it becomes necessary to train and deploy them in a
distributed environment. In an ideal setting, the speed-up obtained in
a distributed setting should be proportional to the number of
computation nodes available in the system.  However in practice
machine learning models often require the computation between the
nodes to be {\em synchronized} resulting in a dramatic reduction in
effective parallelism.

There are different forms of synchronization that can be architected in
a distributed system. The most common form can be described as {\em process synchronization}.
For example consider a shared memory system, where tasks (typically
model parameters updates)  are distributed between
different threads (workers) but there is a common memory bank to which all workers
read and write. Computation is often carried out in phases and at the end of each
phase, all workers wait till the last worker has finished its task and has
saved its computation in shared memory. However, previous studies have shown that
the task time across different workers often follow a skewed distribution and
that the overall time is bottlenecked by the worker which takes the longest
time to finish its task. This is often termed as the Straggler or the Last
Reducer Problem~\cite{cipar2013solving,suri}. Notice that while process synchronization 
ensures that
the output of the computation is sequentially correct, i.e., the same output
is guaranteed to be obtained as if it were executed on a single worker, it
is completely agnostic of the nature of specific task executed.

While process synchronization is problem independent, a new form of
synchronization has emerged specifically for machine learning
tasks. We will refer to it as {\em function
	synchronization}~\cite{li_pserver,Smola2010nips,Garth2013nips}. At
the highest level of abstraction, machine learning reduces to
estimating model parameters with the objective that the model output
will closely align with observable (existing and future)
data. However, depending upon the nature of the specific task, the
{\em loss function} used in the objective function can be different.
Function synchronization relaxes the full process synchronization
barrier by allowing workers to operate in a controlled but
asynchronous manner. For example, workers are allowed to operate using
old outputs of other workers as long as the old values are within a
function-specific bounded delay~\cite{cipar2013solving}. The estimate
of the delay allowed depends upon the nature of the loss
function. However the general rule of thumb is that ``smooth'' loss
functions can tolerate longer delays compared to their non-smooth
counterparts~\cite{li_pserver}. A more radical approach has been
proposed where workers are allowed to update parameters in a
completely asynchronous manner.  When data is extremely sparse (which
is a common occurrence in many application settings) and the
stochastic gradient algorithm is used (thus data access is random),
the chance of update conflicts between workers turns out to be
extremely small.  However the complete asynchronous approach comes
with almost no theoretical guarantees~\cite{recht2011hogwild}.
Nomad~\cite{nomad}, on the other hand, is a non-locking distributed
protocol for matrix completion that leverages function semantics for
ensuring serializability of concurrent updates.

However there is another form of synchronization possible which has
been largely ignored by the machine learning community. We will refer
to it as {\em data-centric synchronization} and has roots in database
transaction systems.  A transaction in a database system is a set of
database operations (typically read, write and update) which are
guaranteed to be executed in an atomic manner.  In order to increase
throughput a modern database systems allows transactions to be
executed in a concurrent fashion while guaranteeing sequential
correctness.  The logic of concurrency in a database system does not
depend upon the semantics of the high level database query but on the
properties of read, write and update operations. Through the use of
carefully designed data access protocols a substantial amount of
concurrency can be achieved in database systems. Recently,
proposals have emerged to leverage optimistic concurrency control (OCC)
from database serializability theory for function synchronization in
the context of machine learning problems such as unsupervised
clustering~\cite{Pan2013nips}.  However we note that the validation step of
OCC in this proposal is used for detection of semantic violation due
to data partitioning and not synchronization violation
due to concurrent data access.

In order to apply data-centric synchronization methods for machine
learning we have to focus on the typical algorithm used to estimate
model parameters rather than the task specific functional form that
describes the model. It may come as a surprise that many machine
learning tasks can be abstracted to carrying out an iterative  
operation based on the template~\cite{bertsekas1989parallel}:
\begin{equation}
	\theta_{i}[\alpha +1] = f_{i}(\theta_{1}[\alpha],\ldots,\theta_{n}[\alpha]) \;\; \forall i=1,\ldots n
\end{equation}\label{mltemplate}
Here each $\theta_{i}$ are the model parameters, $\alpha$ is the iteration number
and $f_{i}$ is the update function for variable $\theta_{i}$ and subsumes the (immutable)
data set. Notice, each $f_{i}$ is only used for updating $\theta_{i}$. We can think of the
parameters as data elements (thus the name parameter database) and a transaction
as a single iteration which consists of updating all the parameters. For simplicity
assume that each $x_{i}$ is assigned to a unique worker and thus, at first
glance,  the worker at iteration $\alpha + 1$ has to {\bf wait for all the parameter values} 
at iteration $\alpha$ to be known, i.e., it has to wait for all other workers
to finish their task.  However by designing access protocols we will
show how the above assumption can be relaxed and transactions (iterations) can
be executed in a concurrent fashion while guaranteeing full sequential correctness.
In the process we will show that the existing database concurrency control
protocols (like
two-phase locking) does not apply in this setting and that  new 
protocols need to be designed to parallelize
fixed point iterative computation  which can overcome the process synchronization barrier.

We summarize our contributions as follows:
\begin{enumerate}
	\item A new form of data-centric synchronization is introduced to speed-up machine learning
	tasks while guaranteeing sequential correctness. 
	New ML systems can be designed which can combine functional and data-centric synchronization as
	they are mutually independent.
	\item
	We will show that traditional data level access protocols like two-phase locking
	are not strong enough
	to support iterative computation which is characteristic of machine learning. 
	\item We will develop a new theory of data-centric synchronization specifically for
	fixed point iterative computation with accompanying provable guarantees for
	sequential correctness.
	\item Experiments on a prototype machine learning task (linear regression) will show that 
	using our relaxed data access protocols we can obtain fifty to eighty percent speed-up 
	compared to implementations that enforce process synchronization.
\end{enumerate}

The rest of the paper is as follows: In Section ~\ref{ml-abstraction} we explain how algorithms for
many ML tasks can be abstracted as a fixed-point iteration computation.
In Section ~\ref{d-cs} and ~\ref{model-db} we present the theoretical foundations of data-centric 
synchronization and relate it to BSP. A simple protocol to implement our proposed
framework is detailed in Section ~\ref{protocol}. Experiments to test the validity of our approach are
presented in Section ~\ref{experiments}. Section ~\ref{delay-section} contains an extension of the data-centric model to incorporate bounded delay updates. We conclude in Section ~\ref{conclusion} with a summary. The supplementary section contains all the proofs.

\section{ML Abstraction and Scope}\label{ml-abstraction}

Machine Learning problems are now being increasingly formulated as 
optimization problems which take a precise form described as
\begin{equation}\label{mleqn}
	\min_{\mathbf{\theta}} \; h(\mathbf{\theta}) \equiv \sum_{j \in \mbox{ \small{data}}}f_{j}(\mathbf{\theta}) + g(\mathbf{\theta})
\end{equation}\label{mlp}
The function $f_{j}$ measures the discrepancy between the model ($\mathbf{\theta}$)
and the data while $g(\mathbf{\theta})$ is a regularizer term to prevent the model from 
overfitting the data and encouraging certain forms of solutions (e.g., sparse or 
spatially contiguous).

The optimization problem as in Equation~\ref{mlp} rarely admit analytical solutions
and recourse is often taken to iterative algorithms like gradient descent which
follow the template of Equation~\ref{mltemplate}. More specifically an update at iteration
$\alpha +1$ is derived from $\alpha$ as
\[
\mathbf{\theta}[\alpha + 1] = \mathbf{\theta}[\alpha] - \eta\nabla h(\mathbf{\theta}[\alpha])
\]
Or expressing it in scalar form
\[
\theta_{i}[\alpha +1] = \theta_{i}[\alpha] - \eta\frac{\partial  h(\mathbf{\theta}[\alpha])}{\partial \theta_{i}}
\]
Note that the component update $\theta_{i}$ depends upon the availability of the full
$\mathbf{\theta}$ component values from the previous iteration.
\subsection{Scope}
A large body of research (both in the ML and optimization community) has tackled problems
related to the convergence of gradient descent and similar methods, setting the
learning parameter $\eta$, the choice of the data term $f_{i}$ and the regularizer
$g$. Our contribution is orthogonal and we will assume that we are operating
in a loss function regime where these issues have been addressed.

Furthermore, the nature of ML solutions is such that the model parameter solution 
can admit a higher degree of imprecision compared to other application domains.
In fact functional synchronization exploits this characteristic of ML solutions.
However, data-centric synchronization will guarantee sequential correctness, i.e.,
we will able to provably show that we can carry certain types of
concurrent (inter-iteration) updates in gradient descent algorithms where
the end result will be exactly like if the algorithm was executed in  a
sequential manner. Extending data-centric syncrhonization to incorporate bounded
delays is relatively straightforward and is briefly explained in the paper.

\section{Data-centric Synchronization}\label{d-cs}
The design space for parallelization can be broadly classified as
follows: \textbf{(i) Data Partitioning:} Training data is divided
among multiple workers and each worker node is responsible for
learning all the parameters based on its chunk of training data;
\textbf{(ii) Feature Partitioning:} A worker node is responsible for
computing updates for a chunk of the feature space based on entire
training data; and \textbf{(iii) Data and Feature
	Partitioning:} A Worker node is responsible for computing updates
for a chunk of the feature space based on a chunk of training
data. Our formal development in this paper is restricted to the
partitioning of the features, i.e., case (iii).

Database Management Systems (DBMSs) significantly simplify the
application development process by providing the \emph{transaction}
abstraction~\cite{GrayTransactionConcept} that makes the issues of
concurrency, synchronization, and failures transparent to the
developer.
Underlying the transaction concept, is a data-centric synchronization
technique such as \emph{two-phase locking}~\cite{EswaranGrayLorieTraiger1976} that synchronizes read
and write accesses from concurrent transactions to ensure that the
interleaved execution of these transactions is correct (formally,
referred to as being serializable~\cite{BeHaGo87}).  When considering
ML computations, a natural question arises if the \emph{transaction}
concept with \emph{two-phase locking} from DBMS can be used for
iterative computations where iterations are parallelized overs
multiple workers ?  A logical mapping will be to view each iteration of
a worker as a \emph{transaction} unit that will ensure that in each
iteration the workers are isolated from each other and are executed
serially. This is not desirable, since the sequential semantics of an
ML computation requires that in each iteration the reads of all workers are
executed before the writes of all workers in the same  iteration.

From the sequential semantics of an ML computation, we observe that
the notion of transaction is not an iteration per worker; rather it
should be mapped to an iteration $\alpha$ across all
workers. Unfortunately, none of the DBMSs to our knowledege support
the notion of splitting a transaction across multiple nodes. Even if
this notion is supported, there is an additional constraint which
requires that the only serialization order of transactions is the one
in which iterations are executed sequentially in strictly increasing
order of the iteration number. To the best of our knowledge, none of
the general-purpose DBMSs support such meta-level synchronization of
transactions.

\section{Model DB Synchronization}\label{model-db}
In this section we develop the theoretical machinery which will
underpin {\em data synchronization}. The key idea is to relax
process synchronization by introducing separate read (RC) 
and write (WC) constraints. We will show that process synchronization
implies RC and WC and that enforcing the two constraints guarantees
sequential correctness while allowing for asynchronous execution.
\subsection{Data Model}
Given the enormous size of the model variable vector, and the fact
that model variables are both read and written during the model
computation, it seems natural to store the model variables as a
database managed over multiple servers.
\begin{defn}
	\textit{\textbf{ML Parameter Database.}} Given a machine learning
	model $M$ with parameters $\Theta =\{\theta_{1},\theta_{2},\ldots,\theta_{m}\},$ a database
	$D$, the parameter database is denoted as $M(\Theta,D)$. 
\end{defn}
Note that in a parameter database $M(X,D)$, the data set $D$ is
immutable and the parameters are inferred from the data using a
machine learning algorithm.  In a shared memory system we are
interested in partitioning the parameters of the machine learning
model so that they can be independently managed and updated by workers.  We
next give a formal definition for partitioning the parameter space
into disjoint partitions that can be managed independently.
\begin{defn}
	\textit{\textbf{ML Feature Partitions.}} A partition set $\Pi$ consists of $p$ partitions over the parameter database $M(\Theta,D)$ denoted by
	\[\Pi = \{\pi_{1}, \pi_{2}, \pi_{3}, \cdots,\pi_{p}\}\]
	where each $\pi_{i}$ may consist of one or more parameters such that:
	(i) $ \pi_{i} = \{\theta_{i_1}, \theta_{i_2}, \theta_{i_3}, \cdots, \theta_{i_d}\}$;
	(ii) $ \forall i,j\ \ \pi_{i}\ \cap\ \pi_{j} = \phi$; and
	(iii) $\Pi = \displaystyle{\bigcup_{i = 1}^p} \pi_{i}$
\end{defn}
\vspace{-0.2cm}
Note that the partitions is a logical concept in that the partitions may be stored in  a single 
database server or in the extreme case may be distributed over $p$ database servers. There will be bijection between workers and partition and 
each worker $i$ will be responsible for updating the partition $\pi_{i}$
associated with it.

\begin{defn}
	\textit{\textbf{Database Access Model.}} An iteration $\alpha$ at worker
	$i$ consists of reads denoted as $r_{i}[\pi_{j}][\alpha]$ and writes $w_{i}[\pi_{i}][\alpha]$. The read and write accesses within each iteration
	$\alpha$ at worker $i$ are such that: $\forall j\ \ r_{i}[\pi_j][\alpha] < w_{i}[\pi_i][\alpha]$,
	where $<$ denotes the \textbf{happens-before} relation. For a system
	consisting of a single worker we will suppress subscripts for the worker
	and denote read and write access as $r[\pi_j][\alpha]$ and
	$w[\pi_j][\alpha]$ respectively.
\end{defn}

\subsection{Sequential ML Execution}
In this section, we illustrate the sequential computation in
Algorithm~\ref{sequential} that will serve as the foundation 
for the correctness semantics of parallelized ML computations.

\begin{algorithm}
	\caption{Sequential Computation}
	\label{sequential}
	\begin{algorithmic}[1]
		\Procedure{ML Computation}{}
		\State initialize iteration $\alpha$
		\While{not converged}
		\State $r[\pi_{1}][\alpha],\ldots,r[\pi_{p}][\alpha]$
		\State fixed point computation at iteration $\alpha$ using data set $D$
		\State $w[\pi_{1}][\alpha],\ldots,w[\pi_{p}][\alpha]$
		\State increment $\alpha$
		\EndWhile
		\EndProcedure
	\end{algorithmic}
\end{algorithm}

Using the read and write model defined above we can now define the
notion of correct executions from an ML computation point-of-view. We use
a single-threaded sequential or synchronous execution as a ground
truth for correctness in an ML system.

\begin{defn}
	\textit{\textbf{Sequential ML Computation.}} A sequential ML
	execution on $\Pi$ is an execution that is single-threaded (i.e., has
	no parallelism across multiple threads) and sequential (i.e., each
	iteration is completed before the next one starts). Formally, in a
	sequential ML execution:
	(i) No operations of an iteration interleaves with the operations of another iteration;
	(ii) Within an iteration, all read operations precede any write operation; and
	(iii) An operation of iteration $\alpha+1$ cannot appear until all operations of $\alpha$ have completed.
\end{defn}
The correctness of sequential execution is based on the observation that if iterations are executed
sequentially and within each iteration all model parameters are read before
being updated then the corresponding execution will preserve the
semantics of an underlying ML computation.

\begin{figure*}[tbph]
	\[{SEQ_1} = r[\pi_1][1]r[\pi_2][1]w[\pi_1][1]w[\pi_2][1]r[\pi_1][2]r[\pi_2][2]w[\pi_1][2]w[\pi_2][2]\]
	\[{SEQ_2} = r[\pi_1][1]r[\pi_2][1]w[\pi_2][1]w[\pi_1][1]r[\pi_2][2]r[\pi_1][2]w[\pi_1][2]w[\pi_2][2]\]
	\caption{Sequential executions over $\Pi = \{\pi_1, \pi_2\}$.\
	}
	\label{1-sequential}
\end{figure*}

Figure~\ref{1-sequential} illustrates executions $SEQ_1$ and
$SEQ_2$ on a model DB consisting of two partitions with two
iterations.  Both executions are sequential ML computation. We note
that within an iteration, the ordering of read and write operations (within
themselves)
can be permuted respectively and the execution will still be deemed
correct.

\subsection{Process-centric Synchronization of Parallel ML Execution}
A sequential ML computation can be parallelized by 
assigning the partitions in $\Pi$ to different worker nodes that execute in
parallel and are each responsible for updating the features in their 
partition. The overall coordination between
the workers is carried out by the master node that initiates
a worker $i$ for each partition $\pi_{i}$ . 
Each worker proceeds (until convergence) in parallel in each iteration
starting with the reading of the current state of $\Pi$,
carrying out the fixed point computation and then updating the value of 
its partition.

Asynchronous execution of workers may lead to race conditions while
reading and writing elements of $\Pi$.  For example, a worker $i$ in
iteration $\alpha+1$ may read stale value of a partition $\pi_j$ if
worker $j$ has not completed its write for the previous iteration
$\alpha$. Similarly, a worker node $i$ in iteration $\alpha$ may
update partition $\pi_i$ before another worker $j$ has had a chance to
read the state of $\pi_i$ for iteration $\alpha$.  Such race
conditions indicate read and write steps of different workers must be
synchronized.  In general, this is accomplished using \emph{process
	synchronization} which is commonly referred to as \emph{bulk
	synchronization (barrier constraints)}.  Algorithm~\ref{barrier-alg}
depicts a parallel ML computation that relies on bulk synchronization
primitives such as \emph{barriers}.

\setcounter{table}{1}
\begin{table}[tbph]
	\begin{subalgorithm}{.5\textwidth}
		\begin{algorithmic}[1]
			\Procedure{Worker $i$}{}
			\State {\bf READ BARRIER}
			\State $r_{i}[\pi_{1}],\ldots,r_{i}[\pi_{p}]$
			\State fixed point computation using data set $D$
			\State {\bf WRITE BARRIER}
			\State $w_{i}[\pi_{i}]$
			\EndProcedure
		\end{algorithmic}
		\caption{Parallel Algorithm with BSP} \label{barrier-alg}
	\end{subalgorithm}
	\begin{subalgorithm}{.5\textwidth}
		\label{barrier-relax}
		\begin{algorithmic}[1]
			\Procedure{Worker $i$}{}
			\State {\bf READ CONSTRAINT}
			\State $r_{i}[\pi_{1}],\ldots,r_{i}[\pi_{p}]$
			\State fixed point computation using data set $D$
			\State {\bf WRITE CONSTRAINT}
			\State $w_{i}[\pi_{i}]$
			\EndProcedure
		\end{algorithmic}
		\caption{Parallel Algorithm with relaxed constraints}
	\end{subalgorithm}
	\captionsetup{labelformat=alglabel}
	\caption{(a) worker computation using read and write barrier and (b) same computation
		with the relaxed read and write constraints.}
	\label{tab:1}
\end{table}

We now express the barrier constraints in the notation of
the database access model introduced in Definition 3.
We express the barrier constraints using logical predicates. In particular,
before the read step at worker $i$ in iteration
$\alpha + 1$, the barrier must ensure that the writes of every worker for
iteration $\alpha$ have completed. This can be stated as:
\[ \emph{\textbf{Read Barrier:}}\;\forall\ i,j,k\ w_{k}[\pi_{k}][\alpha] < r_{i}[\pi_{j}][\alpha+1]\]
Similarly, the write barrier synchronization stipulates that reads of
all workers in iteration $\alpha$ have completed before the worker can
write its partition in iteration $\alpha$. This can be stated as:
\[\emph{\textbf{Write Barrier:}}\;\forall\ i,j,k\ r_{k}[\pi_{j}][\alpha] < w_{i}[\pi_{i}][\alpha]\]

Having formally defined the read and write barriers (and thus
BSP), we show that Algorithm~\ref{barrier-alg} is equivalent to sequential ML
computation as defined in Definition 4 (proof in supplementary section).

\begin{thm}\label{correct-bsp}
	An execution resulting from bulk synchronization (Algorithm 2a) is 
	a sequential ML computation.
\end{thm}

This bulk synchronization of all the workers in each iteration becomes a major inefficiency issue since the results of different worker nodes or threads are not guaranteed to arrive at nearly the same time. At the read barrier, all the workers are blocked until updates from all the workers are completed for the previous iteration. Similarly, at the write barrier, updates of all the workers are blocked until the reads of all the workers are completed at each partition for the current iteration. 

\subsection{Data-centric Synchronization of Parallel ML Execution}
Although the barrier constraints ensures the correctness of ML computations, they impose (unnecessarily) stringent synchronization constraints. 
A key observation is that {\bf  when worker $i$ is reading $\pi_j$, it only needs to be synchronized with respect to the concurrent writes of $\pi_j$}. This is exactly how database accesses from multiple transactions are synchronized. The read synchronization at worker $i$ in iteration $\alpha$ thus can be stated as:
\[\emph{\textbf{Read Constraint:}}\;\forall\ i,j\ w_{j}[\pi_{j}][\alpha] < r_{i}[\pi_{j}][\alpha+1]\]
Similarly, the write synchronization at worker $i$ in iteration $\alpha$ can be stated as:
\[\emph{\textbf{Write Constraint:}}\;\forall\ i,j\ r_{j}[\pi_{i}][\alpha] < w_{i}[\pi_{i}][\alpha]\]

\begin{thm}\label{correct-relax}
	An execution where read and write constraints are enforced (Algorithm 1b) has the same behavior as a sequential ML computation.
\end{thm}

\begin{thm}\label{subsume}
	Executions resulting from bulk synchronization are subsumed in the executions resulting from enforcement of read and write constraints.
\end{thm}

Proofs for Theorem \ref{correct-bsp}, \ref{correct-relax} and \ref{subsume} are given in Appendix 1. A direct consequence of theorem \ref{subsume} is that there are less constraints (more possible executions) leading to greater concurrency without compromising on the 
correctness (Theorem \ref{correct-relax}).

\section{A Data-centric Synchronization Protocol}\label{protocol}
We present a simple protocol for data-centric synchronization to ensure correctness of ML computations. \emph{Worker processes} run independently of each other and can be either on the same or different machines. A central server process is responsible for communicating the convergence of the algorithm to the workers so that
they can stop their work. Workers send their read and write requests to the server and the
computation is performed locally over the full training data set. The server is responsible for executing read and write operations on the data objects while ensuring
that the read and write constraints are enforced on the parameters.

{\sf The Write Protocol.}
A write operation issued from a worker $i$, which is in its $\alpha^{th}$ iteration, on parameter partition chunk $\pi_i$ can be executed if this chunk has been read by all the worker processes in their $\alpha^{th}$ iterations. This can be ensured in a very efficient way by associating a bit vector (of size equal to number of workers) with each parameter chunk. When a chunk is updated, all bits in this vector are set to zero. When a read operation issued by a worker $k$, while in its $\alpha^{th}$ iteration, is executed on this chunk, the bit corresponding to this worker is set. The scheduler can execute the above mentioned write operation by quickly checking if all bits in the bit vector are one. Otherwise, the write operation is deferred for later consideration.

{\sf The Read Protocol.}
A read operation issued from a worker $i$, which is in its $(\alpha+1)^{th}$ iteration, on a parameter chunk $j$ can be executed if a write operation issued from a worker $j$, while in its $\alpha^{th}$ iteration, has already been executed on this chunk. Again, this can also be ensured in a very efficient and simple way by associating an iteration number corresponding to each chunk. Every time a write operation is executed on this chunk, its iteration number is set to iteration number of the write operation. The above mentioned read operation can be executed on this chunk if the iteration number in the read operation is one more than the iteration number of the chunk. Otherwise, the read operation is deferred for later consideration.

We note that the protocol outlined above is very different from the lock-based
synchronization that is used in general purpose database systems. The main reason
being that data-centric syncrhonization for iterative computation requires
different approaches than traditional methods for enforcing serializability~\cite{GrayTransactionConcept}.

\section{Experiments}\label{experiments}
In order to test our approach for practical applicability, we
conducted an exhaustive evaluation (Figure~\ref{exp}) using both synthetic and a real
world dataset.  All experiments were performed on a machine with
dual Intel(R) Xeon(R) CPU E5-2697 v2 @ 2.70GHz CPUs. Each of these
CPUs have 24 cores (48 threads total). The machine has 256 GB memory.
We next describe various experiments and discuss the results. The
results were obtained by running the experiments 10 times each and
taking trimmed mean (average after dropping 2 fastest and 2 slowest
runs). While in practice distributed ML solutions will be deployed to handle
large data, the impact of data-centric synchronization can be observed
even on relatively small data sets.

\subsection{Scaling with Number of Workers}\label{sec-workers}
\begin{figure}[bpht]
	\begin{subfigure}{0.3\textwidth}
		\centering
		\includegraphics[clip,trim=2cm 1cm 2.4cm 1cm,width=1\textwidth]{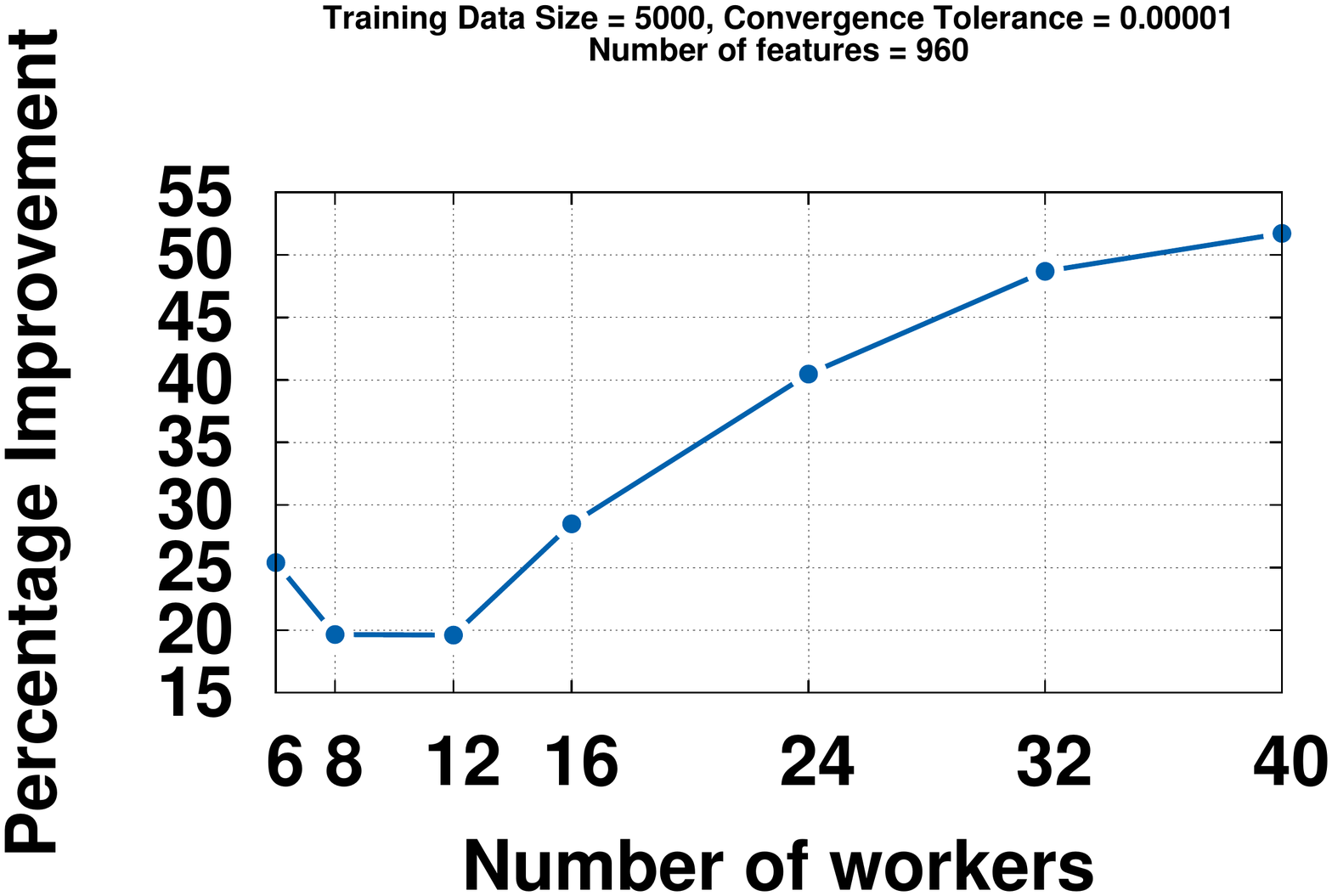}
		\caption{Scaling with number of workers (Batch GD)}
		\label{improvement}
	\end{subfigure}
	\begin{subfigure}{0.3\textwidth}
		\centering
		\includegraphics[clip,trim=2.3cm 1cm 4cm 1cm,width=1\textwidth]{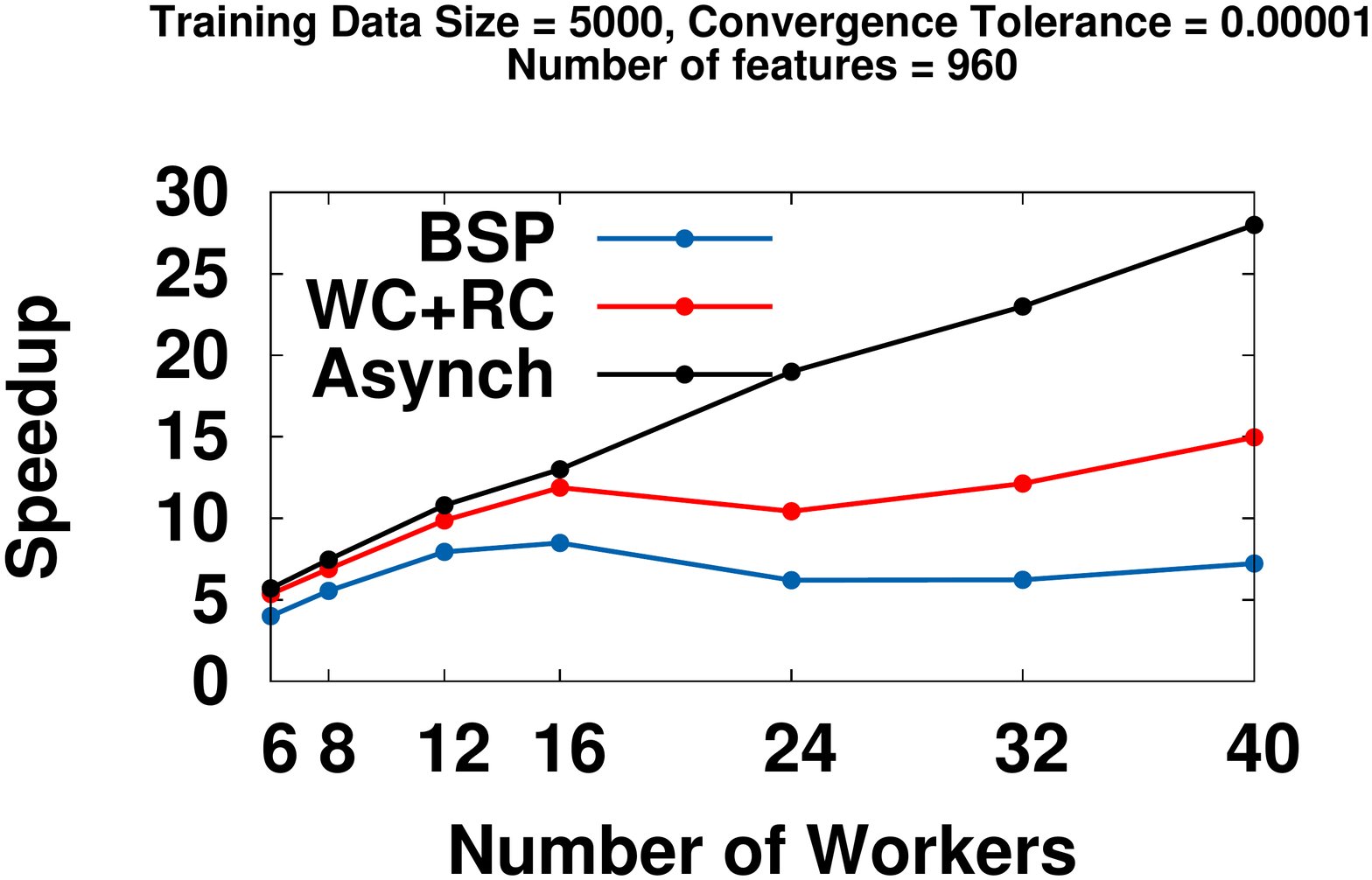}
		\caption{Speedup (Batch GD)}
		\label{speedup}
	\end{subfigure}
	\begin{subfigure}{0.3\textwidth}
		\centering
		\includegraphics[clip,trim=2cm 1cm 2.2cm 1cm,width=1\textwidth]{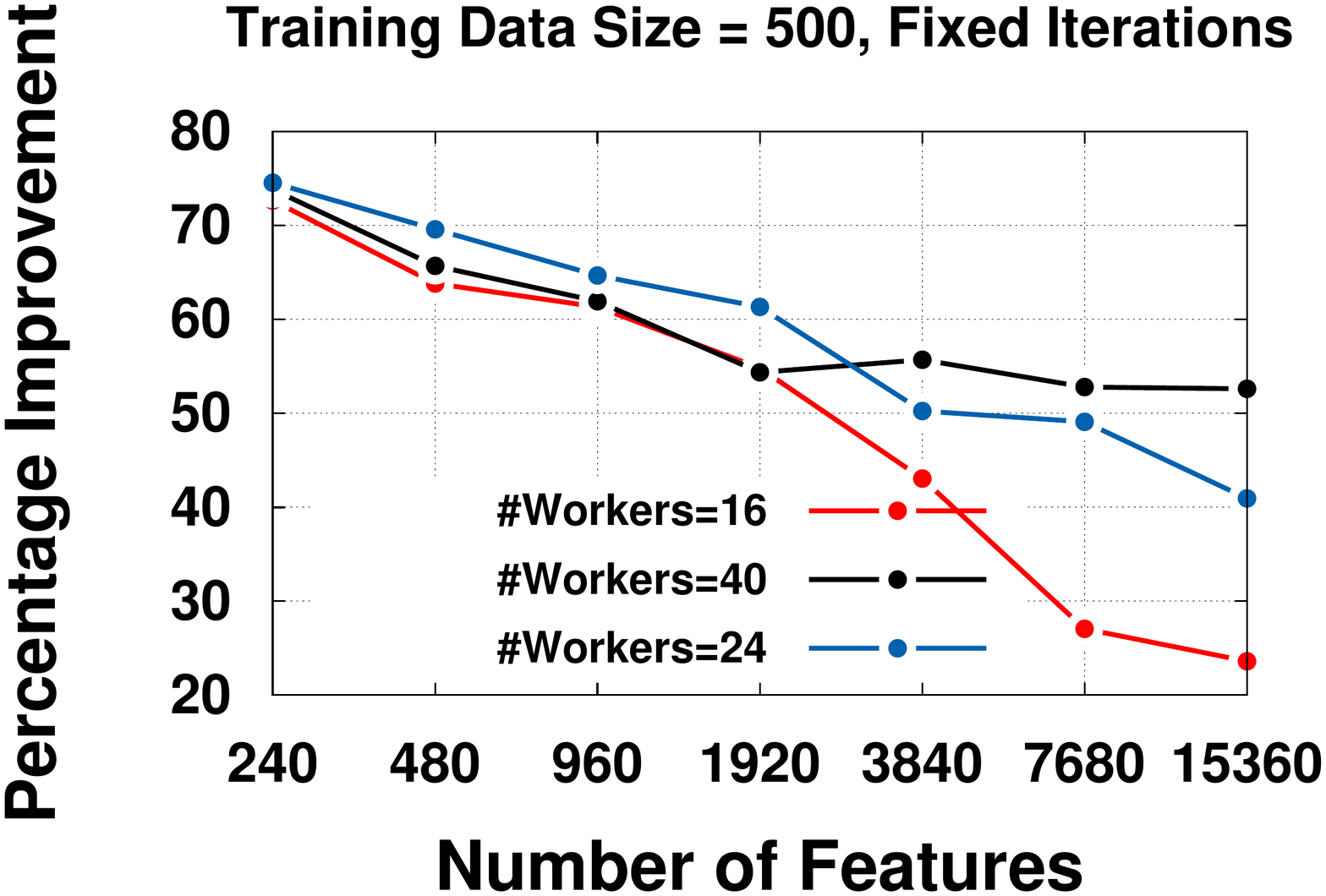}
		\caption{Scaling with number of features (Batch GD)}
		\label{features}
	\end{subfigure}
	\begin{subfigure}{0.32\textwidth}
		\centering
		\includegraphics[clip,trim=2cm 1cm 3cm 1cm,width=1\textwidth]{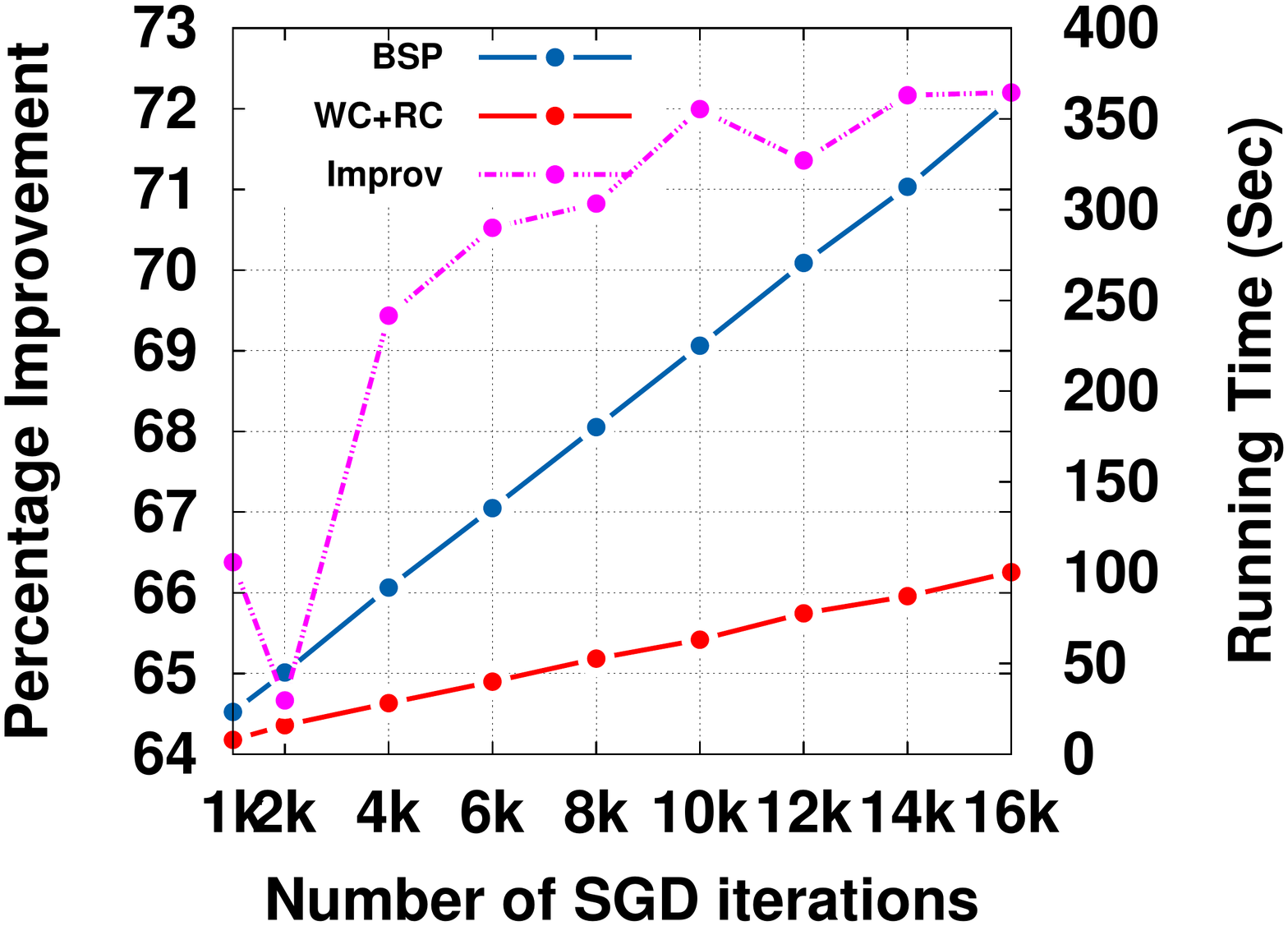}
		\caption{Performance with SGD iterations on real dataset}
		\label{real}
	\end{subfigure}
	\begin{subfigure}{0.32\textwidth}
		\centering
		\includegraphics[clip,trim=1.5cm 1cm 2.5cm 1cm,width=1\textwidth]{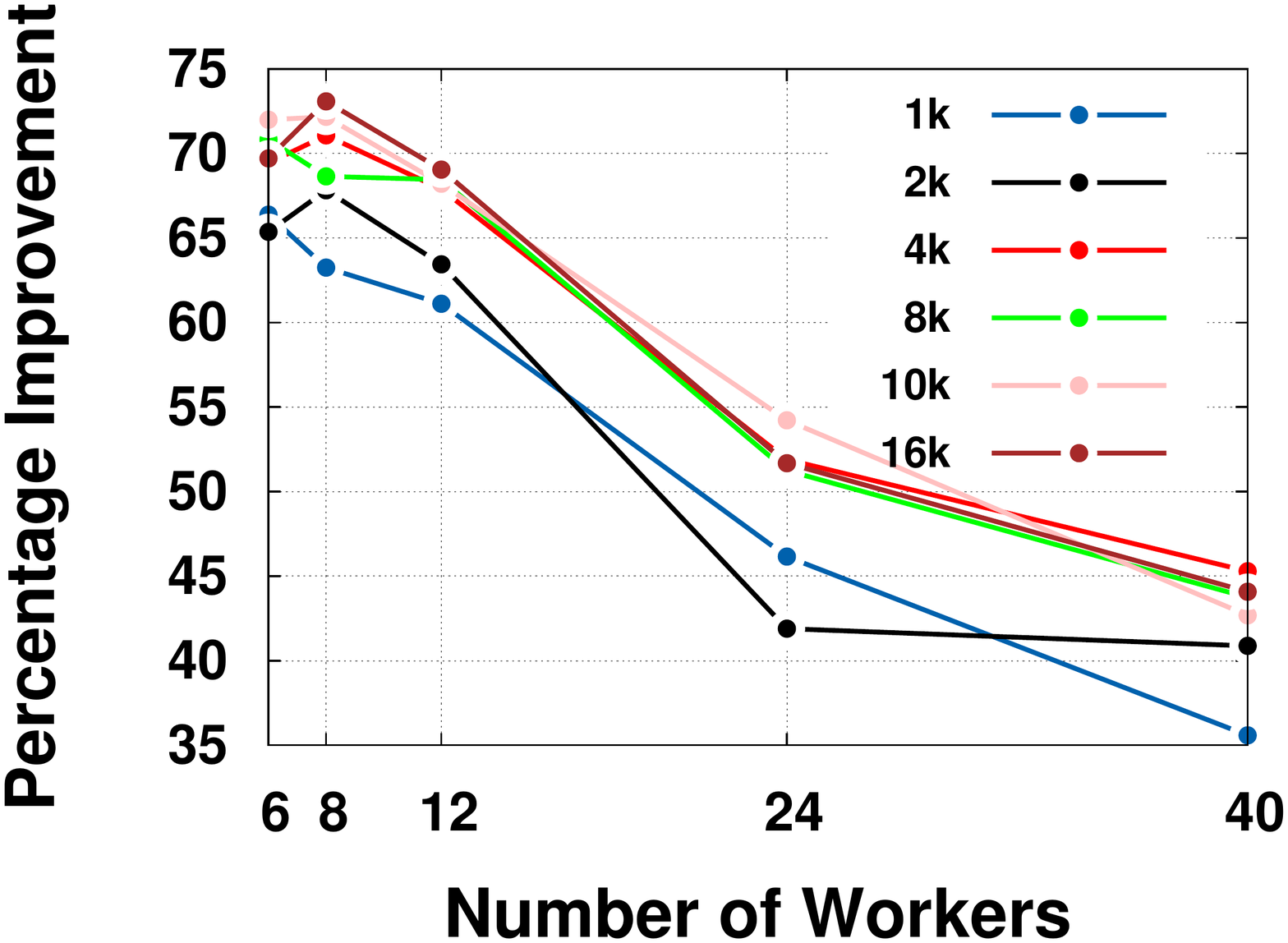}
		\caption{Performance with SGD iterations on real dataset with varying parallelism}
		\label{real-vary}
	\end{subfigure}
	\begin{subfigure}{0.32\textwidth}
		\centering
		\includegraphics[clip,trim=1.5cm 1cm 2.5cm 1cm,width=1\textwidth]{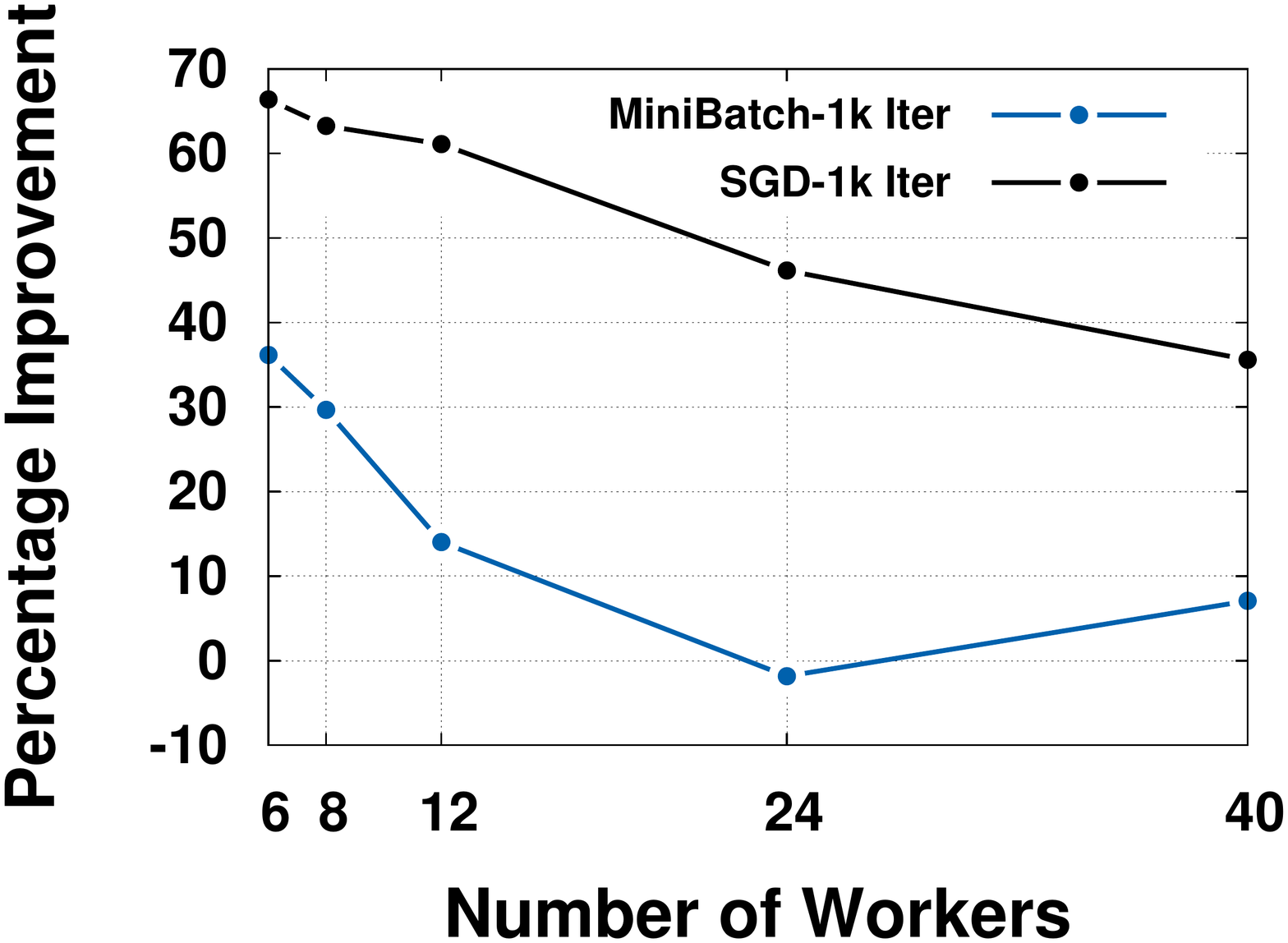}
		\caption{Performance with SGD and mini-batch iterations on real dataset}
		\label{real-mini}
	\end{subfigure}
	\caption{Experimental Results\label{exp}}
\end{figure}

For this experiment, we generated a synthetic dataset with 960
numerical features and one dependent variable. A linear regression
model was trained over 5000 examples (gradient descent iterations
until convergence). Number of workers was varied from 6 to
40. Figure~\ref{improvement} shows the percentage improvement (from 20\% to almost 55\%) in the running time of the training phase. As the
number of workers increases, data-centric synchronization gets more opportunity for
improvement over process-centric synchronization due to the wait for more workers to
finish read and write operations in each iteration. 
Figure~\ref{speedup} illustrates the speedup under the two approaches and we observe that 
under BSP the speedup is relatively flat whereas data-synchronization achieves significantly better speedup. We have also included a curve for the theoretical limit for a completely asynchronous
speed up based on Amdhal's law using 0.01 for the fraction of the computation that is neccessarily serial due to memory contention and related issues. 

\subsection{Scaling with Number of Features}
For this experiment, we generated several synthetic datasets each with
different number of features. Linear regression models were trained on
500 examples for each of the datasets (constant number of gradient
descent iterations). All the experiments were run with number of
workers set to 16, 24 and 40. Figure~\ref{features} shows the
scalability of our approach for large number of features.  With less
number of features, the trend of improvement obtained with different
number of workers is not clear but there is a clear trend as the
number of features are increased (with larger number of workers, we
get more improvement). In particular, with 16 workers, the percentage
improvement declines significantly from the high of 75\% to 25\% as the
number of features is increased. However, when the amount of
parallelism is increased by deploying more workers, the overall
improvement with a large number of features is around 50\% indicating
that data-centric synchronization does result in higher-level of
parallelism and concurrency in the system.
\subsection{Experiments with real world dataset}
In Figures~\ref{real},~\ref{real-vary},~and~\ref{real-mini}, we
compare our approach with BSP on a real world
dataset~\cite{kogan2009predicting} which has 150,360 features and
16,087 training examples. Figure~\ref{real} reports the performance
improvement and absolute times of running stochastic gradient descent
(SGD) with varying number of iterations with 6 workers. On real dataset,
our approach results in significant improvements since the percentage improvement ranges between 65\%
to almost 75\%. Figure~\ref{real-vary} reports the performance
improvements for different number of iterations by varying the number of
workers. In this experiment, we get a consistent trend that the
percentage improvement declines from the high of 70-75\% to 40-50\%
(which is still a significant improvement). The explanation of this
decline is that under SGD the amount of
work done per iteration is much smaller (1 training sample versus the entire dataset). Furthermore, as the
number of workers is increased from 6 to 40, the work per assignment
gets further reduced due to feature partitioning. This results in synchronization overhead
becoming dominant over useful computation. By analyzing the raw data, we discovered that the
rate of increase of the synchronization overhead of BSP is relatively
less than that with our technique. This can be explained that as the
amount of useful work being done by each worker becomes smaller, the
window of delay between slowest  and the fastest worker finishing
their respective assignment becomes increasing smaller. On the other
hand, under our scheme the synchronization check with a larger number
of workers increases while the amount of useful work per unit
continues to decrease. This explains the overall decline in percentage improvement with increased parallelism.
In order to validate this hypothesis, in
Figure~\ref{real-mini} we compared the relative performance of the two
protocols computing Gradient Descent with a mini-batch where the batch size was fixed to 100
training examples. In relative terms, indeed the decline in
performance improvement is much more pronounced in SGD whereas it is
not as sharp under mini-batch. Detailed analysis of raw data indicates
that in the case of mini-batch, both approaches benefit from increased
parallelism but beyond a certain point increased parallelism is not
beneficial. We attribute this to a small amount of useful computation
per iteration. This performance degradation can be addressed by
exploiting data partitioning and data sparsity. 

\section{Data-centric Synchronization with Admissible Delay}\label{delay-section}
Function-based synchronization exploit the semantics of the underlying
function being optimized and leverage it to increase asynchrony
(equivalently, reduce synchronization and permit more concurrency and
interleavings among workers) in the parallel ML computation. 
This asynchrony can be captured by allowing the read operations to
read stale writes. Similarly, a write operation in an iteration does
not have to be blocked until the reads from all the workers in that
iteration have been processed. Instead, the write of a worker can be
processed as long as all the reads are within some distance of the
write (where distance is specified in terms of the iteration number).
The protocol has a notion of delay $\delta \geq 0$ which stipulates that
workers are separated from each other (i.e., read and write
operations) within $\delta$ iterations.

The weaker form of read synchronization at worker $i$ in iteration
$\alpha$ thus can be stated as:
\[\emph{\textbf{Asynchronous Read Constraint:}}\;\forall\ i,j\ w_{j}[\pi_{j}][\alpha - 1 - \delta] < r_{i}[\pi_{j}][\alpha]\]
Similarly, the weaker form of write synchronization at worker $i$ in iteration $\alpha$ can be stated as:
\[\emph{\textbf{Asynchronous Write Constraint:}}\;\forall\ i,j\ r_{j}[\pi_{i}][\alpha - \delta] < w_{i}[\pi_{i}][\alpha]\]
We note that when $\delta = 0$, the above constraints yield executions that ensure sequential semantics. On the other hand, if 
$\delta = \infty$, the parallel execution is completely asynchronous and reduce to parallel executions resulting from a system such as Hogwild!~\cite{recht2011hogwild}.

\subsection{Revised Protocol}
{\sf The Write Protocol.}
A write operation issued from a worker $i$, which is in its $\alpha^{th}$ iteration, on parameter partition chunk $\pi_i$ can be executed if the slowest worker to read this chunk is no more than $\delta$ iterations behind worker $i$. To ensure this, an array (of size equal to number of workers) can be associated with each parameter chunk. When a read operation issued by worker $k$, while in its $\beta^{th}$ iteration, is executed on this chunk, the element corresponding to this worker is set to $\beta$. The scheduler can execute the above mentioned write operation by checking if the minimum number in this array is greater than or equal to $\alpha - \delta$. Otherwise, the write operation is deferred for later consideration.

{\sf The Read Protocol.}
A read operation issued from a worker $i$, which is in its $\alpha^{th}$ iteration, on a parameter chunk $j$ can be executed if the last write operation executed on this chunk was issued by worker $j$, while in its $\alpha - \delta - 1^{th}$ iteration or later. To ensure this, an iteration number can be associated with each chunk. Every time a write operation is executed on this chunk, its iteration number is set to the iteration number during which this write operation was issued. The above mentioned read operation can be executed on this chunk if the iteration number of the chunk is greater than or equal to ($\alpha - \delta - 1$). Otherwise, the read operation is deferred for later consideration.

\section{Conclusion}\label{conclusion}
In this paper  we have presented a new data-centric synchronization paradigm for carrying
out machine learning tasks in a distributed environment.  
Our approach abstracts the iterative nature of ML algorithms and
introduces specific read and write constraints (RC and WC)  whose enforcement 
guarantees sequential correctness while providing opportunity to speed up ML computation. We also show that the bulk synchronization process (BSP) design pattern, which is extensively
used in distributed ML tasks, implies RC and WC.  Our proposal complements function
synchronization techniques in  distributed ML research which uses ``bounded staleness'' to relax BSP and increase throughput.

\bibliographystyle{abbrv}
\bibliography{nips-bib}
\clearpage
\section{Proofs}
\begin{thm}\label{correct-BSP}
	An execution resulting from bulk synchronization (Algorithm 1a) is a sequential ML computation.
\end{thm}

{\em Proof.} Let ${BSP}$ be an execution using 
partition set $\Pi$.
involving $p$ partitions and $p$ workers as shown
in Algorithm 1a. 
In order to show that ${BSP}$ is sequential, 
we need to establish:
\begin{itemize}
	\item No operations of two different iterations are interleaved.
	\item Within each iteration, all reads precedes any write.
	\item Iterations are executed consecutively.
\end{itemize}
\textbf{Condition 1} follows from the read barrier which enforces that writes of the
previous iteration are completed before reads from the next iteration
can begin.

\textbf{Condition 2} is a consequence of the write barrier.

\textbf{Condition 3} is a consequence of the fact that 

\[\forall k, w_{k}[\pi_{k}][\alpha] < w_{k}[\pi_{k}][\alpha + 1]\] (combining read and write barrier).
\hfill$\Box$

\begin{thm}\label{correct-relax}
	An execution where read and write constraints are enforced (Algorithm 1b) has the same behavior as a sequential ML computation.
\end{thm}
{\em Proof:} We need to ensure that following conditions are satisfied for every partition:
\begin{itemize}
	\item No operation on a partition in an iteration interleaves with the operations on the same partition in another iteration.
	\item Within an iteration, all read operations on a partition precede any write operation on the same chunk.
	\item An operation on a partition in iteration $\alpha+1$ cannot appear until all operations on the same partition in $\alpha$ have completed.
\end{itemize}
To prove \textbf{condition 1}, let's assume that an execution contains a fragment $op_{1}[\pi_{i}][\alpha]op_{2}[\pi_{j}][\alpha+1]op_{3}[\pi_{k}][\alpha]$, where $op$ is either read or write operation. We have to show that $j\neq k$. We can prove this by contradiction by starting with assumption $j=k$. The following cases are possible

i) $op_{2}=R$ and $op_{3}=R$

This means that for some $x$ and $y$

\begin{equation}\tag{1}\label{1}
r_{x}[\pi_{j}][\alpha+1]<r_{y}[\pi_{j}][\alpha]
\end{equation}

However, from read constraint it follows that

\begin{equation}\tag{2}\label{2}
w_{j}[\pi_{j}][\alpha]<r_{x}[\pi_{j}][\alpha+1]
\end{equation}

From \ref{1} and \ref{2}, 

\[w_{j}[\pi_{j}][\alpha]<r_{y}[\pi_{j}][\alpha]\]

This is a violation of our write constraint.

ii) $op_{2}=R$ and $op_{3}=W$

This means that for some $x$
\[r_{x}[\pi_{j}][\alpha+1]<w_{j}[\pi_{j}][\alpha]\]

This case is a direct violation of read constraint.

iii) $op_{2}=W$ and $op_{3}=R$

This means that for some $x$

\begin{equation}\tag{3}\label{3}
w_{j}[\pi_{j}][\alpha+1]<r_{x}[\pi_{j}][\alpha]
\end{equation}

However, from write constraint, it follows that

\begin{equation}\tag{4}\label{4}
r_{j}[\pi_{j}][\alpha+1]<w_{j}[\pi_{j}][\alpha+1]
\end{equation}

and from read constraint, it follows that 

\begin{equation}\tag{5}\label{5}
w_{j}[\pi_{j}][\alpha]<r_{j}[\pi_{j}][\alpha+1]
\end{equation}

From \ref{3}, \ref{4} and \ref{5}, it follows that $w_{j}[\pi_{j}][\alpha]<r_{x}[\pi_{j}][\alpha]$, which violates write constraint.

iv) $op_{2}=W$ and $op_{3}=W$

This means that
\begin{equation}\tag{6}\label{6}
w_{j}[\pi_{j}][\alpha+1]<w_{j}[\pi_{j}][\alpha]
\end{equation}

However, from write constraint, it follows that
\begin{equation}\tag{7}\label{7}
\forall\ i, r_{i}[\pi_{j}][\alpha+1]<w_{j}[\pi_{j}][\alpha+1]
\end{equation}

and from read constraint, it follows that
\begin{equation}\tag{8}\label{8}
\forall\ i, w_{j}[\pi_{j}][\alpha]<r_{i}[\pi_{j}][\alpha+1]
\end{equation}

From \ref{7} and \ref{8}, it must be the case that 
\[w_{j}[\pi_{j}][\alpha]<w_{j}[\pi_{j}][\alpha+1]\] which is exactly opposite to \ref{6}.

\textbf{Condition 2} directly follows from the write constraint.

\textbf{Condition 3} follows from read and write constraints. Read constraint stipulates that any read on a chunk in iteration $\alpha+1$ happens only after the write on this chunk in iteration $\alpha$ has finished. And write constraint stipulates that the write on this chunk in iteration $\alpha$ happens only after all reads on this chunk in iteration $\alpha$ are finished. Thus, all operations in $\alpha$ are finished before any operation in $\alpha+1$ begin. 

\begin{thm}\label{subsume}
	Executions resulting from bulk synchronization are subsumed in the executions resulting from enforcement of read and write constraints.
\end{thm}

{\em Proof:} It is clear from the proof given in Theorem~\ref{correct-relax} that relaxed constraints are special cases of barrier conditions applied on per partition level.

\subsection{Examples of Executions}
In Figure~\ref{history-examples}, we show three possible executions on a model database.  Consider a model database with two partitions $\{\pi_1, \pi_2\}$. Assume that
the ML computation consisted of two iterations. Execution $H_1$ is the one that results from BSP and from Theorem~\ref{correct-BSP}, it obviously results in correct results. Execution $H_2$ is one of the several more exeuctions possible by relaxing the barrier conditions in our model. In Theorem~\ref{correct-relax}, we showed that these executions also give the correct results. However, $H_3$ is an example of executions that are permitted neither by the BSP nor the RC and WC. These executions lead to incorrect results. More possible executions as compared to BSP lead to increased concurrency and hence performance improvement.

\begin{figure}[H]
	\[H_1 = r_1[\pi_1][1]r_1[\pi_2][1]r_2[\pi_1][1]r_2[\pi_2][1]w_1[\pi_1][1]w_2[\pi_2][1]r_1[\pi_1][2]r_1[\pi_2][2]r_2[\pi_1][2]r_2[\pi_2][2]w_1[\pi_1][2]w_2[\pi_2][2]\]
	\[H_2 = r_1[\pi_1][1]r_1[\pi_2][1]r_2[\pi_1][1]r_2[\pi_2][1]w_2[\pi_2][1]r_1[\pi_2][2]w_1[\pi_1][1]r_1[\pi_1][2]r_2[\pi_1][2]r_2[\pi_2][2]w_1[\pi_1][2]w_2[\pi_2][2]\]
	\[H_3 = r_1[\pi_1][1]r_1[\pi_2][1]w_1[\pi_1][1]r_2[\pi_1][1]r_2[\pi_2][1]w_2[\pi_2][1]r_1[\pi_1][2]r_1[\pi_2][2]w_1[\pi_1][2]r_2[\pi_1][2]r_2[\pi_2][2]w_2[\pi_2][2]\]
	\caption{Examples of execution histories of two worker computations with two iterations over $\{\pi_1, \pi_2\}$.}
	\label{history-examples}
\end{figure}

\end{document}